\documentclass[11pt,a4paper]{article}
\usepackage[makeroom]{cancel}
\usepackage{bm}
\usepackage{amsmath}
\usepackage{amsfonts}
\usepackage{graphicx}
\usepackage{nicefrac}
\usepackage{authblk}
\usepackage{cite}
\usepackage[left=2.2cm,top=2.2cm,right=2.2cm,bottom=2.2cm,nohead]{geometry}
\DeclareGraphicsExtensions{.png,.jpg,.pdf}
\usepackage{epstopdf}
\usepackage{float}
\usepackage{subfigure}
\usepackage{fouriernc} 
\usepackage{enumitem}
\setlist{nolistsep,leftmargin=*}

\title{Capacitive Deionization -- defining \\ a class of desalination technologies}

\usepackage{authblk}
\makeatletter
\renewcommand\AB@authnote[1]{\textsuperscript{\normalfont#1}}

\makeatother

\author[1]{P.M.~Biesheuvel,}
\author[2]{M.Z.~Bazant,}
\author[3]{R.D.~Cusick,}
\author[2]{T.A.~Hatton,}
\author[4]{K.B.~Hatzell,}
\author[5]{M.C.~Hatzell,}
\author[6]{P.~Liang,}
\author[7]{S.~Lin,}
\author[8]{S.~Porada,}
\author[9]{J.G.~Santiago,}
\author[10]{K.C.~Smith,}
\author[11]{M.~Stadermann,}
\author[2]{X.~Su,}
\author[12]{X.~Sun,}
\author[13]{T.D.~Waite,}
\author[14]{A.~van~der~Wal,}
\author[15]{J.~Yoon,}
\author[16]{R.~Zhao,}
\author[17]{L.~Zou,}
\author[18]{M.E.~Suss}

\affil[1]{Wetsus, European Centre of Excellence for Sustainable Water Technology, 
The Netherlands}
\affil[2]{Department of Chemical Engineering,  Massachusetts Institute of Technology, USA}
\affil[3]{Department of Civil \& Environmental Engineering,
University of Illinois at Urbana-Champaign, USA}
\affil[4]{Department of Mechanical Engineering, Vanderbilt University, USA}
\affil[5]{School of Mechanical Engineering, Georgia Institute of Technology, USA}
\affil[6]{School of Environment, Tsinghua University, Beijing, China}
\affil[7]{Department of Civil and Environmental Engineering, Vanderbilt University, USA}
\affil[8]{Faculty of Science and Technology, 
University of Twente, The Netherlands} 
\affil[9]{Department of Mechanical Engineering, Stanford University Stanford, USA}
\affil[10]{Department of Mechanical Science and Engineering, University of Illinois at Urbana-Champaign, USA}
\affil[11]{Lawrence Livermore National Laboratory, USA}
\affil[12]{EST Water \& Technologies, Beijing, China}
\affil[13]{School of Civil and Environmental Engineering, University of New South Wales, Sydney, Australia}
\affil[14]{Department of Environmental Technology, Wageningen University, The Netherlands}
\affil[15]{School of Chemical and Biological Engineering, Seoul National University, Republic of Korea}
\affil[16]{School of Physics and Materials Science, East China Normal University, Shanghai, China}
\affil[17]{Masdar Institute, Khalifa University of Science and Technology, 
United Arab Emirates}
\affil[18]{Faculty of Mechanical Engineering, Technion -- Israel Institute of Technology, Israel}

\date{} 

\begin{document}


\renewcommand{\t}{\widetilde}
\renewcommand{\t}{}
\newcommand{\s}[1]{\mathrm{_{#1}}}

\maketitle

\begin{abstract}
Over the past decade, capacitive deionization (CDI) has realized a surge in attention in the field of water desalination and can now be considered as an important technology class, along with reverse osmosis and electrodialysis. While many of the recently developed technologies no longer use a mechanism that follows the strict definition of the term ``capacitive'', these methods nevertheless share many common elements that encourage treating them with similar metrics and analyses. Specifically, they all involve electrically driven removal of ions from a feed stream, storage in an electrode (i.e., ion electrosorption) and release, in charge/discharge cycles. Grouping all these methods in the technology class of CDI makes it possible to treat evolving new technologies in standardized terms and compare them to other technologies in the same class. 

\end{abstract}

\begin{center}{\noindent\rule{12cm}{0.4pt}}\end{center}
 
\newpage

Capacitive Deionization (CDI) is a method of water desalination using electrodes that, in a cyclic fashion, are electronically charged and discharged with resultant adsorption and release of salt. There has been growing interest in CDI over the past decade with the technology now attracting considerable industrial and academic attention. Currently, about one hundred publications annually deal with the topic of CDI and this number is expected to increase in coming years. 

While the concept of water desalination using porous electrodes involving a cyclic process of salt removal and release dates back to the 1960s, the term ``capacitive deionization'' is of much more recent origin with the first use in the public record in a 1995 conference proceedings by Farmer \textit{et al.}~\cite{Farmer_1995}. Historically, the term CDI has
been used to describe desalination with carbon electrodes by the mechanism of electrical double layer (EDL) formation.  

However, other mechanisms for water desalination, following the same strategy of cyclic operation combined with charge input, have recently been explored. These mechanisms include water desalination by redox-active intercalation materials, ion-selective molecules or polymers immobilized at an electrode surface that are regenerated through applied voltages, as well as systems that are hybrids of multiple mechanisms (e.g., intercalation and EDL). With the emergence of these different mechanisms for ion storage in porous electrodes, the question has arisen as to whether these other mechanisms can also be categorized as CDI. In this document, we advocate that the terminology of CDI should be used to define a broad class of technologies which share the same methods of analysis and metrics. We propose that CDI includes all methods of water desalination that share the same general features of process operation (namely, the cyclic nature of ion storage and release driven by electrical charging of electrodes) and which share the same set of relevant metrics, including equilibrium salt removal and stored charge as a function of cell voltage. These operational and analytical aspects define whether a certain technology or chemistry can be categorized as CDI and recognized as a technology platform for water desalination alongside reverse osmosis and electrodialysis.
The grouping of the different mechanisms within the technology class of CDI underscores the fact that common analytical methods and operational strategies apply to each of these technologies, and will facilitate quantitative comparison between the methodologies, which will be supportive to the development of the field as a whole.

Thus, it is our proposal that CDI is used to describe a class of desalination technologies where electrodes are charged and discharged in a cyclic manner, regardless of the exact removal mechanism or electrode material. This implies that the various materials and methods used to this effect all fall within the class of CDI, including processes involving the use of carbon with and without chemical modifications, such as with bound ion-selective molecules, and also the use of intercalation materials (for instance based on sodium manganese oxide, or transition metal hexacyanoferrates), in processes based on redox or Faradaic reactions (and described as either a battery or pseudocapacitive material). While most work in CDI is with an electrode fixed in position, often in the form of a thin layer or film, CDI is also possible with moving electrodes, either as rods or wires that can be switched between solutions, or as particles, acting as electrodes, that flow through the cell. As in fixed electrode systems, these mobile electrodes are cyclically charged and discharged (in the frame of reference of the moving electrode) with resultant adsorption and release of salt.

Water desalination by CDI can have various objectives including the removal of most of the salt from water (as sometimes implied by the word deionization) or the removal of only a portion of the salts from any water source, be it brackish water or seawater. Finally, and of increasing importance, is selective ion removal by the use of double layer effects, high affinity chemical groups, redox-active materials and membranes, with these selective separations equally included as a possible objective of CDI technologies.


\begin{thebibliography}{99}

\bibitem{Farmer_1995}
J.C. Farmer, D.V. Fix, G.V. Mack, R.W. Pekala, and J.F. Poco, ``The Use of Capacitive Deionization with Carbon Aerogel
Electrodes to Remove Inorganic Contaminants from Water,'' Low Level Waste Conference, Orlando, USA (1995).

\end{thebibliography}
\end{document}